# Impact of Land Use on the DOM Composition in Different Seasons in a Subtropical River Flowing through the Region Undergoing Rapid Urbanization

Author: Qi Liu [a, b], Yuan Jiang [a, b, *], Zhaojiang Hou [a], Yulu Tian [a], Kejian He [c], Lan Fu [d], and Hui Xu [e]

[a] Beijing Key Laboratory of Traditional Chinese Medicine Protection and Utilization, Faculty of Geographical Science, Beijing Normal University, Beijing 100875, China

[b] State Key Laboratory of Earth Surface Processes and Resource Ecology, Faculty of Geographical Science, Beijing Normal University, Beijing 100875, China

[c] College of Resource and Environment, Yunnan University, Kunming 650091, China.

[d] Shenzhen Academy of Environmental Sciences, Shenzhen 518001, China

[e] Department of Biostatistics and Epidemiology School of Public Health and Health Sciences, University of Massachusetts Amherst, Amherst MA 01003, USA

**ABSTRACT:** The dissolved organic matter (DOM) composition in river ecosystems could reflect the human impacts on the river ecosystem, and plays an important role in the carbon cycling process. We collected water and phytoplankton samples at 107 sites in the Dongjiang River in two seasons to assess the impact of the sub-catchments land use structure on the DOM composition. The results showed that (1) the forested sub-catchments had higher humic-like C1 (16.45%) and C2 (25.04%) and lower protein-like C3 (22.57%) and C4 (35.95%) than urbanized and mixed forest-agriculture sub-catchments, while the urbanized sub-catchments showed an inverse trend (4.54%, 15.51%, 33.97% and 45.98%, respectively). (2) The significant variation in the proportion of C1 and C4 between the dry and rainy seasons was recorded in both the forested and the mixed forest-agriculture sub-catchments ($p<0.01$), but only C4 showed an obvious seasonal variation in the urbanized sub-catchments ($p<0.01$). While the DOM composition was mainly related to the proportion of urbanized land and forested land year-round ($p<0.01$), it had stronger correlation with forested land in the dry season and urbanized land in the rainy season. (3) No significant correlation between the DOM composition and the agricultural land proportion was found in either season ($p>0.05$). Our findings indicated that the DOM composition was strongly dependent on the land use structure of the sub-catchments and varied seasonally, but the seasonal variation pattern could be disturbed by human activities in the extensively urbanized catchments. Notably, the higher C4 proportion compared with those in temperate rivers implied a larger amount of $CO_2$ released from the subtropical rivers into the atmosphere when considering bioavailability.

**Key words:** Dissolved organic matter; Land use structure; Seasonal variation; Subtropical river; Carbon cycle

## 1. Introduction

The dissolved organic matter (DOM), mainly composed of humic-like and protein-like materials, can be found globally in freshwater ecosystems (Hosen et al., 2014; Yang et al., 2016). Previous studies have evidenced that, among the DOM compositions, certain humic-like substances are generally derived from the streams in impervious landscapes, such as forest and wetlands (Vidon et al., 2014; Wilson et al., 2016). Meanwhile, other kinds of humic-like materials also originate from wastewater, especially those components found in agricultural catchments (Fuß et al., 2017; Tye and Lapworth, 2016). The protein-like components, such as tryptophan-like and tyrosine-like materials, are related to sewage inputs and/or microbial activity in streams (Meng et al. 2013). Usually, tryptophan-like materials are detected at a relative large proportion in urban rivers or streams polluted by urban sewage (Borisover et al., 2011; Yu et al., 2016; Zhou et al., 2016), while tyrosine-like materials are thought to be a product of algae and autotrophic microorganism (Korak et al., 2015) and are generally found in the urban rivers, into which sewage rich in nutrients and microbial activities has been discharged (Liang et al., 2015). Regarding the difference in sources, the DOM composition in a terrestrial river ecosystem could demonstrate both the water quality and land use structure of the drainage basins. Changes in land use structure could be, therefore, recognized through the variations in the DOM composition, which in turn reflect the human impacts on the river ecosystem via land use changes.

The DOM composition is not simply an indicator of the anthropogenic impacts on river ecosystems, it also plays an important role in the process of carbon cycling by microbial communities processing in freshwater ecosystems (Balch and Guéguen, 2015; Shin et al., 2016; Williams et al., 2016; Wilson et al., 2016). Based on the chemical components, protein-like materials have a lower C:N ratio and are generally considered to be more bioavailable than the humic-like materials (Kaplan and Bott, 1989; Shin et al., 2016). Increases in the abundance of bioavailable protein-like materials may stimulate microbial metabolism. Microbial production of enzymes for acquiring the metabolic carbon depends on its composition and the availability of other organic and inorganic forms of nitrogen and phosphorus. The humic-like materials usually have multiple aromatic rings and can play an important non-consumptive role in metabolism by acting as an electron shuttle in redox reactions. Therefore, changes in the DOM composition also alter ecosystem functions, such as respiration and inorganic nutrient cycling (Parr et al., 2015). For a similar reason, whether organic carbon from headstreams is delivered downstream, enters the river food webs, or is metabolized and released from rivers in a gaseous form could be, to a large extent, dependent on the DOM composition. In this sense, the effects of the human impacts on land use could be transmitted to ocean ecosystems through changes to the DOM composition in rivers.

Considering the importance of the DOM in global biochemical cycling, the relationship between the DOM composition and human activities has become an important topic (Catalán et al., 2013; Fuß et al., 2017; Yu et al., 2015). Increasingly more researchers have focused on interpreting this relationship in different areas (Fellman et al., 2008; Masese et al., 2017; Williams et al., 2013). William et al. (2016) found a significant high protein-like proportion when the anthropogenic intensity increased. Hosen et al. (2014) showed a rise in the proportion of protein-like material along with increasing anthropogenic intensity. Parr et al. (2015) found that humic-like materials significantly decreased as the anthropogenic intensity increased. These studies recognized the DOM composition of rivers in temperate regions and the related impacts from different factors. We selected a series of sub-catchments with different land use combinations in the Dongjiang River Basin, a subtropical river basin in southern China that has undergone rapid urbanization, and we aimed to recognize the characteristics of the DOM in subtropical river ecosystems and its relationship to intensive land use changes.

The Dongjiang River watershed is located in a subtropical humid monsoon region of China, and this region is famous for its rapid industrialization and urbanization over the last four decades. Shenzhen, located in its downstream plain, has developed in approximately 30 years from a small fishing village into a big city with a population of more than 20 million (Fu et al., 2015). Now, the downstream plain of the Dongjiang River Basin has become an important part of the Pearl River Delta Metropolis. We hypothesize that, first, the composition of the DOM in the Dongjiang River should be strongly dependent on the land use structure of the sub-catchments within the watershed, and that second, in comparison to those in the temperate regions, the composition of the DOM in subtropical river should be characterized by relative high proportion of protein-like materials, considering the high microorganism activity under the tropical climatic condition. Third, the composition of the DOM should vary among the seasons due to its response to the subtropical monsoon climate.

## 2. Materials & methods

### 2.1. Study area

The Dongjiang River, which is mainly located in the Guangdong Province in southeastern China, is one of three main tributaries of the Pearl River, with a drainage area of approximately 35,340 km$^2$ (Fig. 1). The Dongjiang River Basin belongs to a subtropical monsoon climate with a mean annual temperature of 21 ℃. There are two distinct seasons, the rainy season from April to October and the dry season from November to March of the following year. The mean temperature of the rainy and dry season is 26.3 ℃ and 15.1 ℃, respectively. The annual precipitation is 1800-1900 mm and the mean annual runoff is approximately $296 \times 10^8$ m$^3$, over 80% of which occurs in the rainy season (Ding et al. 2016; Lv et al. 2013). In the whole watershed, the evident land use differences were presented from the forested area in the upstream to the urbanized area in the downstream along the Dongjiang River Basin (Liao et al., 2013).

**2.2. Sampling for water.**

Samples for water and phytoplankton were collected at 107 sites along the main stem and major tributaries of the Dongjiang River during the rainy season (August in 2015) and dry season (December in 2015) (Fig. 1). There was no significant rain precipitation (<10 mm over 24 h) during the sampling process.

Water and phytoplankton samples in total were collected 0.5 m below the water surface using a 2.5-L organic glass hydrophore. The sampling sites were located in the middle of the stream or at least 5 m from the riverbank. Samples were stored in 1 L polyethylene bottles. Water samples were filtered using Whatman GF/F 0.45 μm filters that were precombusted at 450 ℃. The water samples were stored in a refrigerator box (< 4 ℃) and transported to the laboratory in a frozen state prior to further processing. Phytoplankton samples were preserved in Lugol's iodine solution. In the laboratory, phytoplankton samples were left undisturbed for more than 24 h to mature, and were concentrated to approximately 30 mL (National Environmental Protection Bureau, 2002).

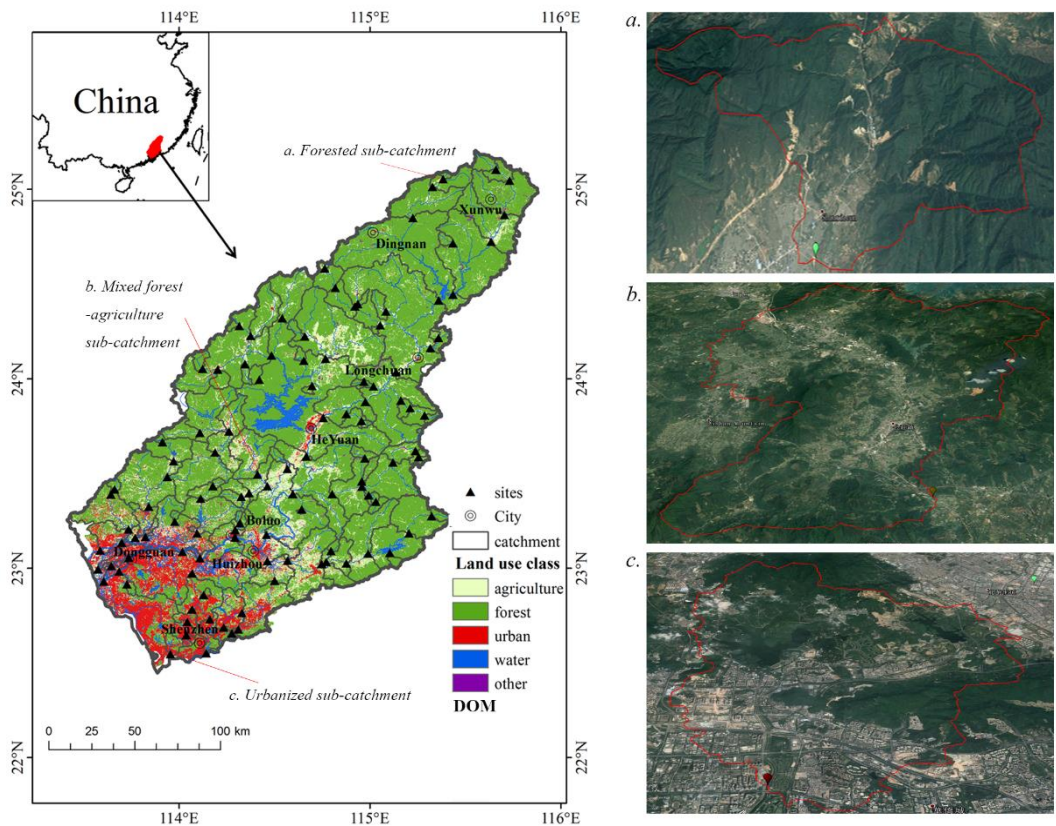

**Fig. 1 - Distribution of 107 sampling sites and location in the Dongjiang River Basin.**

## 2.3. Measuring of stream water samples

DOM fluorescence excitation−emission matrices (EEM) were measured using an Itachi-7000 spectrofluorometer following established methods (Osburn et al., 2016; Stedmon and Bro, 2008). The EEM spectra of samples and Milli-Q water (blank EEMs) were scanned from 200 nm to 450 nm at 5 nm increments, and the emissions were measured at each excitation wavelength from 300 nm to 600 nm at 2 nm increments. The scan rate for all samples was set to 1200 nm/min. Each sample was analyzed in triplicate. The sample EEMs were obtained by subtracting the water blank EEM.

PARAFAC analysis was used to obtain the fluorescence components of the DOM by the DOMFluor Toolbox in Matlab 2013b (Stedmon and Bro, 2008). The database was made of 642 EEMs in the Dongjiang River. Four components were identified using the PARAFAC model, which was validated by split-half analysis (Yu et al., 2015), and each component was identified based on the shape and location of the components observed in previous studies (Table 1) (Coble et al., 1990; Hosen et al., 2014). Components 1 and 2 (C1 and C2) were humic-like materials, which were similar to a humic-like fluorophore of terrestrial origin and an anthropogenic humic-like fluorophore, respectively. Components 3 and 4 (C3 and C4) were protein-like materials, which were similar to tryptophan-like and tyrosine-like materials. The proportion of each component was calculated by the fluorescence intensity of each component divided by the sum of the fluorescence intensity of all components.

Table 1 - Emission and Excitation Maxima and Characteristics of the PARAFAC Components

| Components | Excitation (nm) | Emissions (nm) | Characteristics |
| --- | --- | --- | --- |
| C1 | <250 (330) | 400-440 | humic-like, terrestrial origin |
| C2 | <250(285) | 340-380 | humic-like，anthropogenic fluorophore |
| C3 | 220-230(280) | 320-336 | protein-like, tryptophan-like |
| C4 | <250(290) | 336-350 | protein-like, tyrosine-like |

The permanganate index ($COD_{Mn}$, mg/L) was determined by the potassium permanganate method. Total nitrogen (TN, mg/L), total phosphorus (TP, mg/L) and chlorophyll a (chl-a, μg/L) were measured using a spectrophotometer. In the laboratory, the cell abundance of phytoplankton (CP) was directly counted in a plankton counting chamber (0.1 mL) using a Nikon ECLIPSE 90i optical digital microscope at an eyepiece magnification of l0× and an objective magnification of 40×. The cell numbers of phytoplankton were determined in 100 random fields (National Environmental Protection Bureau, 2002; Hu, 1980). In this study, we used these parameters to represent the effect factors of the DOM composition.

**2.4. Classification and statistics of the land use**

Land use in the Dongjiang River Basin was mapped by Landsat Thematic Mapper (TM) satellite imagery with 30 m resolution in 2009. The land use types were classified into five groups base on Ren et. al: forested, agricultural, and urban land use, water bodies and other land use with the ERDAS 9.2 (Fig. 1) (Ren et al., 2011).

A 30 m Digital Elevation Model (DEM) was used to determine the flow directions by the hydrology tool in ArcGis 10.0. The sub-catchment for each site was produced using the sampling site as the outlet point. Notably, the sub-catchment for a lower stream site encompasses the sub-catchment corresponding to any upper stream site (Ding et al. 2016; Donohue et al. 2006). Then the proportion of forested (Pfor, %), agricultural (Pagr, %) and urbanized (Purb, %) land uses in each sub-catchment area was calculated in this study.

**2.5. Data analysis.**

The principal component analysis (PCA) was used to group the 107 sub-catchments based on their land use data. The paired t-test or Wilcoxon test was conducted to explore the seasonal variation of the DOM composition or water chemical parameters. The regression was used to analyze the impact of land use structure on the DOM composition. The Redundancy analysis (RDA) was used to determine the relationship between the DOM composition and water chemical and biological parameters. The variables that were not normally distributed by the Kolmogorov–Smirnov test (Feio et al., 2009) were log-transformed from the variable plus one prior to analysis.

## 3. Results

## 3.1. Characteristics of the DOM composition for sub-catchment groups with different land use structures

From the results of the PCA (Fig. 2), 107 sub-catchments were divided into three groups according to their land use data. The sub-catchments of the first group (green sites) were more than 85% covered by forested land. The sub-catchments of the second group (red sites) had over 20% urbanized land, very little agricultural land, and a large area of forested land. In addition, the sub-catchments of the third group (orange sites) had a mixed land use composition predominantly by forested and agricultural land. We named three groups as the forested sub-catchment group, urbanized sub-catchment group and mixed forest-agriculture sub-catchment group, respectively.

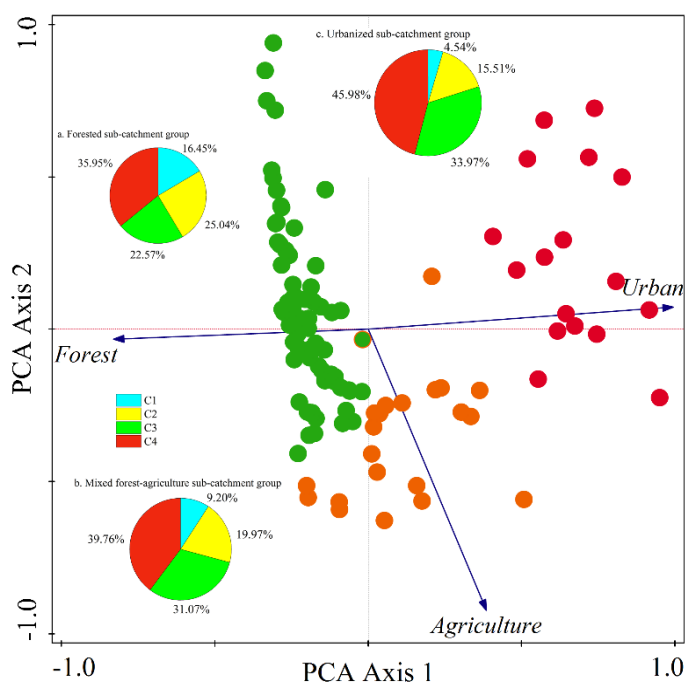

**Fig. 2 - The PCA results of 107 sub-catchments according to their land use data. The green, red and orange sites represent the forested sub-catchment group, urbanized sub-catchment group and mixed forest-agriculture sub-catchment group, respectively.**

The streams in the forested sub-catchment group had the highest average proportion of humic-like C1 (16.45%) and C2 (25.04%) and the lowest average proportion of protein-like C3 (22.57%) and C4 (35.95%) among all three groups (Fig. 2). In contrast, the streams in the urbanized sub-catchment group had the lowest average proportion of humic-like C1 and C2 (4.54% and 15.51%, respectively), and the highest average proportion of protein-like C3 and C4 (33.97% and 45.98%, respectively). The streams in the mixed forest-agriculture sub-catchment group had average proportions of C1, C2, C3 and C4 between those of the forested and urbanized sub-catchment groups (9.20%, 19.97%, 31.07% and 39.76%, respectively).

### 3.2. The DOM compositions in the dry and rainy seasons

The results of paired t-test or Wilcoxon test showed that there were significant seasonal variations in C1 and C4 ($p<0.05$) in streams of the forested sub-catchment group (Fig. 3). The humic-like C1 was higher in the rainy season, whereas the protein-like C4 was higher in the dry season. In streams of the mixed forest-agriculture sub-catchment group, the humic-like C1 and C2 was higher in the rainy season ($p<0.05$), whereas the protein-like C3 and C4 was higher in the dry season ($p<0.05$). In streams of the urbanized sub-catchment group, only C4 showed an evident seasonal variation, which had a higher contribution in the rainy season ($p<0.01$).

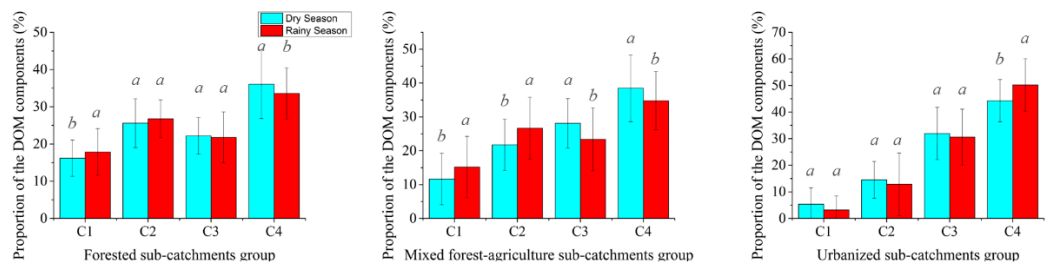

**Fig. 3 - The DOM compositions in the dry and rainy season under different sub-catchment groups. a and b represent the variation between seasons.**

## 3.3. Quantitative relationship between the DOM composition and land use

The results of the regression showed that (Fig. 4), the proportion of C1, C2, C3 and C4 was significantly correlated with the proportion of forested land and urbanized land whether in the dry and rainy season ($p<0.001$). The DOM composition was mainly related to the forested land proportion in the dry season ($p<0.01$). The proportion of humic-like C1 and C2 increased, and the protein-like C3 and C4 decreased, as the proportion of forested land use increased. In the rainy season, the DOM composition was the most closely correlated to the urbanized land proportion ($p<0.001$). The proportions of humic-like C1 and C2 were negatively correlated with those of the urbanized land, while the proportions of protein-like C3 and C4 were positively correlated. No significant correlation between the DOM composition and the agricultural land proportion was found in either season ($p>0.05$).

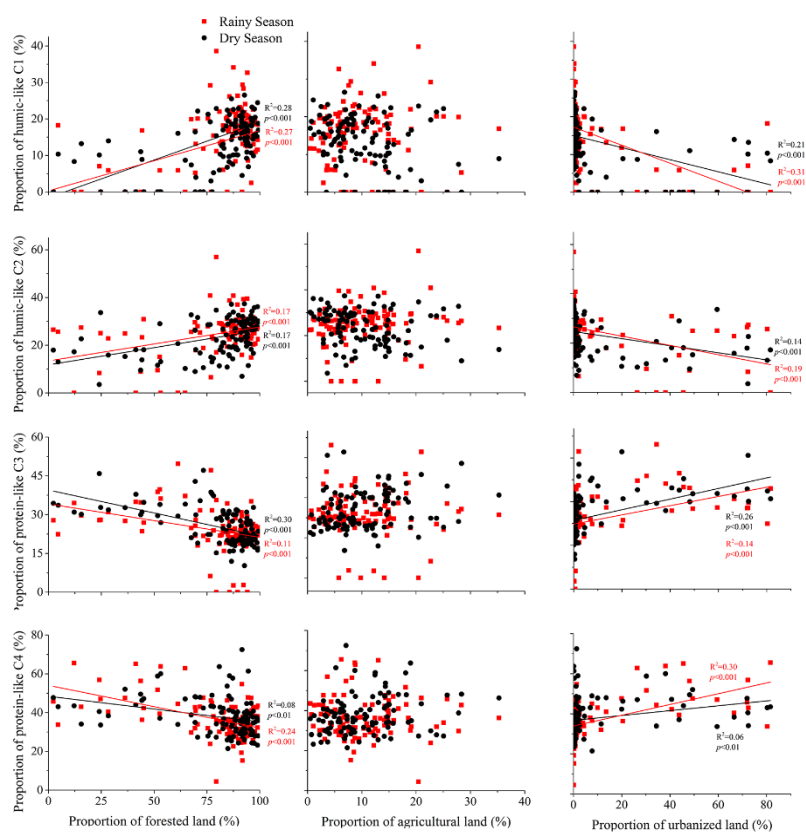

Fig. 4 - The regression models based on the DOM composition and land use data

## 3.4. The water chemical and biological parameters and their relationships to the DOM compositions in the dry and rainy seasons.

According to the results of the paired t-test or Wilcoxon test, $COD_{Mn}$, and the CP had higher concentration in the rainy season ($p<0.05$) in the forested sub-catchment group (Table 2), and chlorophyll a, TN and TP showed no seasonal variation ($p>0.05$). In the mixed forest-agriculture sub-catchment group, the concentration of chlorophyll a was higher in the dry season ($p<0.05$), and the CP and $COD_{Mn}$ were higher in the rainy season ($p<0.05$), while the concentration of TN and TP had no seasonal variation ($p>0.05$). In the urbanized sub-catchment group, the biological parameters had a higher concentration in the rainy season ($p<0.05$). However, the water chemical parameters ($COD_{Mn}$, TN and TP) showed weak variations across different seasons ($p>0.05$).

We used RDA to analyze the relationship between the proportion of components in the total DOM and water chemical and biological parameters (Fig. 5). The results showed that all parameters had a significant explanation ratio in both seasons ($p<0.01$), which was 18.6% and 30.3%, respectively. In both seasons, we found that the proportions of humic-like C1 and C2 had significant downward trends as all water chemical parameters increased ($p<0.01$). The proportions of protein-like C3 and C4 grew with the concentration of the water chemical parameters increasing ($p<0.01$). The proportion of protein-like C4 had a significant increasing trend as the biological parameters rose ($p<0.01$).

Table 2 - Seasonal Characteristics of Water Chemical and Biological Parameters in Different Sub-catchment Groups

| | the forested group | the forest-agriculture mixed group | the urbanized group |
|---|---|---|---|

|  | Dry season | Rainy season | Dry season | Rainy season | Dry season | Rainy season |
| --- | --- | --- | --- | --- | --- | --- |
| Chl-a (µg/L) | 4.52±4.50 a | 4.38±5.03 a | 8.26±4.25 a | 5.06±11.68 b | 6.69±5.47 b | 7.01±35.11 a |
| CP (×10$^4$ cells/L) | 6.90±21.68 b | 14.55±38.32 a | 19.65±63.86 b | 106.3±252.66 a | 41.04±53.72 b | 119.48±796.46 a |
| COD$_{Mn}$ (mg/L) | 1.81±0.94 b | 2.68±1.29 a | 2.46±1.15 b | 3.10±1.18 a | 5.60±1.40 a | 5.53±1.71 a |
| TP (mg/L) | 0.06±0.11 a | 0.06±0.12 a | 0.09±0.08 a | 0.08±0.07 a | 0.46±0.92 a | 0.56±0.92 a |
| TN (mg/L) | 0.97±1.36 a | 1.03±1.20 a | 1.77±4.86 a | 1.68±1.78 a | 10.02±6.21 a | 8.02±4.95 a |

Chla-a represents the concentration of the chlorophyll a. CP represents the cell abundance of phytoplankton. a and b represent the variation between seasons.

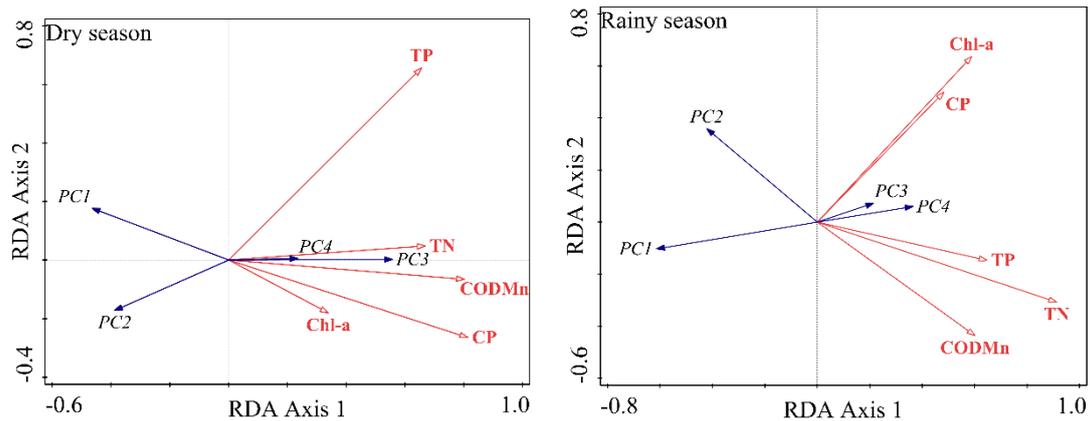

**Fig. 5 - The relationships between the DOM compositions and water chemical and biological parameters in the dry and rainy season. PC1, PC2, PC3 and PC4 indicate the proportions of C1, C2, C3 and C4 in the total DOM, respectively.**

## 4. Discussion

### 4.1. Response of the DOM composition to the land use structure

According to our results from the stepwise regression, the forested land and the urbanized areas were the principal contributors to the DOM composition in streams. The impact of agricultural land on the DOM composition was not determined in the Dongjiang River Basin. Whether these results occur as a general pattern in various regions, or exist as a particular feature of our study area should be evidenced by further studies over a greater region. However, the results could be reasonably interpreted as a general pattern given the limited agricultural land use area in the studied catchment. The percentage of agricultural land area did not exceed 36% in any of the sub-catchments involved in this study. Under these conditions, the impact of agricultural land on the DOM composition, particularly on the proportion of C2, which is generally believed to relate to agricultural land use (Graeber et al., 2012; Xu et al., 2017), seemed to be disturbed by the influence of the urbanized area, since a negative correlation between the proportion of C2 and the percentage of urbanized area was evidently recorded (Fig. 4). The results from the RDA (Fig. 5) support this disturbance further in that the proportion of C2 correlated negatively to the concentration of $COD_{Mn}$ and the concentrations of the nutrients originating mainly from the urban sewage.

**4.2. High proportion of protein-like material C4 in streams of forest-dominated landscape**

The composition of the DOM in the Dongjiang River demonstrated the characteristics related not only to the land use structure but also to other factors. These could be demonstrated by the high proportion of C4. Based on the data in the previous studies, we estimated that the proportion of C4 was 15.1% on average in the Great Lakes Region (Williams et al., 2016), and 14% in central and southern Maine (Parr et al., 2015). In the present study, the proportion of C4 was as high as 35.95% even in streams of sub-catchments dominated by forests (Fig. 2).

We speculate that one reason for such a high proportion of C4 could be the thermal condition of the river ecosystem. The Dongjiang River Basin is located in a humid subtropical region of China, and the mean annual air temperature is $21^{\circ}C$, which is almost 5 times higher than in the temperate region where relatively low proportions of C4 were recorded (Parr et al., 2015; Williams et al., 2016). The high temperature in our study area may accelerate phytoplankton reproduction and, therefore, cause an increase in the proportion of C4 (Korak et al., 2015; Rhee and Gotham, 1981). Another reason for such a high proportion of C4 may be due to orange and litchi orchards that are distributed throughout our study area. The organic fertilizer and pesticide used in orchard management might be carried to the river by erosion and runoff. These chemicals could lead to a direct increase in both the proportion of C4 (Osburn et al., 2016) and the concentration of nutrients, such as TN and TP. The latter may further enhance the proportion of C4 by boosting phytoplankton reproduction. In consideration of the bioavailability, an increase in the proportion of C4 implies an enhanced microbial process, which may lead to the release of more $CO_2$ from the inland river ecosystems directly into the atmosphere, decreasing the amount of carbon carried to the ocean.

**4.3. Impact of urbanization on the seasonal variation of the DOM composition**.

The rainy season of the Dongjiang River Basin spans from April to October. It is characterized by high monthly mean air temperature and a high total precipitation, compared to the dry season from November to March of the following year. The DOM composition of the Dongjiang River may respond to the seasonal climate conditions. The results of our study did show some seasonal variations in the proportions of different DOM components (Fig. 3). The variation in the proportion of C1 between the two seasons was recorded in streams of both the forested and the mixed forest-agriculture sub-catchment groups, which might imply an intrinsic intra-annual fluctuation pattern in the DOM composition in the Dongjiang River under limited anthropogenic effects. The seasonally varied proportion of C4, recorded in streams of the forested sub-catchment group, could be interpreted as being influenced by the intrinsic seasonal variation in runoff related to seasonal precipitation (Catalán et al., 2013; Oliver et al., 2016). The sub-catchments covered predominantly by forested land were usually located upstream, where the runoff quantity should be limited and was not regulated by reservoirs. The protein-like C4 derived mainly from wastewater discharge could be concentrated in the dry season due to a reduction in the quantity of runoff, therefore, leading to an increase in its proportion.

However, the seasonal variations in the stream DOM composition could be changed obviously in the urbanized sub-catchment group. The proportions of C1, C2 and C3 did not show any significant differences between the two seasons, and the proportion of C4 had the opposite trend compared to the forested sub-catchment group. These results suggest that the intrinsic seasonal variation pattern in the DOM composition could be greatly disrupted by human activities in extensively urbanized catchments, which still may not attract enough attention for the evaluation of the impacts of urbanization on carbon-related biochemical cycling in subtropical rivers.

## 5. Conclusion

The results from this study have provided significant evidence of the impacts of both land use and climatic conditions on the stream DOM composition and carbon flux between inland freshwater ecosystems and the atmosphere. One of the important findings from this work was the recognition on a notably high proportion of protein-like materials in streams. The high proportion of those materials resulted from not only the decrease of forested land area, but also the large quantities of discharges from upper surface runoff and wastewater that could in turn boost phytoplankton reproduction. The other interesting finding was the seasonal variation in proportions of some stream DOM components. This variation was mainly characterised as a higher proportion of protein-like materials in the dry season than that in the rainy season, which could be, however, changed in the sub-catchments dominated by urban area. These findings suggested that the urbanization might to a large extent increase $CO_2$ released from streams into the atmosphere in subtropical rivers than that in temperate rivers, considering the higher proportion of the bioavailable protein-like materials under the warmer climate condition. Urbanization may also alter the seasonal rhythm of carbon cycling via disrupting the seasonal variation of the streams in the proportion of protein-like materials.

**\*** Corresponding author. Beijing Key Laboratory of Traditional Chinese Medicine Protection and Utilization, Faculty of Geographical Science, Beijing Normal University, Beijing 100875, China. Tel: +86 010 58806093. E-mail address: jiangy@bnu.edu.cn.

**Acknowledgements**

We are grateful to Feipeng Ren, Xing Xiong, Qiuzhi Peng, Jianyu Liao, Leting Lv and Yan Wen for their help in the field work. This work was supported by the National Science and Technology Major Project: Water Pollution Control and Management Technology of China (No. 2017ZX07301-001-003). We would like to thank in advance the anonymous reviewers for their contributions.